\documentclass[twoside]{JHEP3}
\usepackage{amsmath}
\newcommand{\li}{\mbox{\small{{\it li}}}\,} 
\newcommand{\dsr}{\mbox{\small{{\it  dsr}}}\,}
\newcommand{\skp}{S_{\mbox{\small{{\it kp}}}}\,}

\title{Particle    and    Antiparticle    sectors    in    DSR1    and
$\kappa$-Minkowski space-time}

\author{R.~Aloisio\\
INFN, Laboratori Nazionali del Gran Sasso, 
67010 Assergi, L'Aquila (Italy)\\
E-mail: \email{aloisio@lngs.infn.it}}
\author{J.M.~Carmona \\ 
Departamento de F\'{\i}sica Te\'orica, Universidad de
Zaragoza, C/ Pedro Cerbuna 12, E-50009 Zaragoza (Spain) \\
E-mail: \email{jcarmona@unizar.es}}
\author{J.L.~Cort\'es \\ 
Departamento de F\'{\i}sica Te\'orica, Universidad de
Zaragoza, C/ Pedro Cerbuna 12, E-50009 Zaragoza (Spain) \\
E-mail: \email{cortes@unizar.es}}
\author{A.~Galante \\
INFN, Laboratori Nazionali del Gran Sasso, 
67010 Assergi, L'Aquila (Italy) and
Dipartimento di Fisica dell'Universit\`a di L'Aquila,
67100 L'Aquila (Italy)\\
E-mail: \email{galante@lngs.infn.it}} 
\author{A.F.~Grillo\\
INFN, Laboratori Nazionali del Gran Sasso, 
67010 Assergi, L'Aquila (Italy)\\
E-mail: \email{grillo@lngs.infn.it}}
\author{F.~M\'endez\\
INFN, Laboratori Nazionali del Gran Sasso, 
67010 Assergi, L'Aquila (Italy)\\
E-mail: \email{mendez@lngs.infn.it}}

\abstract{
In this paper we explore the problem of antiparticles in DSR1 and
$\kappa$-Minkowski space-time following three different approaches
inspired by the Lorentz invariant case: a) the dispersion relation, b)
the Dirac equation in space-time and c) the Dirac equation in 
momentum space. We find that it is possible to
define a map $S_{\dsr}$ which  gives the antiparticle sector from the
negative frequency solutions of the wave equation. 
In $\kappa$-Poincar\'e, the corresponding map $\skp$
is the antipodal mapping, which is different from $S_{\dsr}$. 
The difference is related to the composition law, which is
crucial to define the multiparticle sector of the theory. 
This discussion permits to show that the energy of the antiparticle 
in DSR is the positive root of the dispersion relation, which is
consistent with phenomenological approaches.
}


\begin{document}

\section{Introduction}
The idea that Lorentz symmetry could be modified at very high
energies, owing to quantum gravity effects, has been intensively
explored in the last years~\cite{lorqg}. For example in string
theory, the possibility that space-time could have a non-commutative
structure~\cite{connes} opens the door to a Lorentz non-invariant
world~\cite{carroll}; in loop quantum gravity, on the other hand,
neutrinos and photons might not satisfy a Lorentz invariant dispersion
relation~\cite{alfaro}.

The fact that such violation of Lorentz invariance (LI) happens at
certain distance (or energy) scale\footnote{The main candidate is the
Planck scale.} and  that this scale should be the same for any
observer,  led  to the  main  idea of  the  so-called 
Doubly Special Relativity (DSR) principle,
developed by G.~Amelino-Camelia~\cite{amelino1,rev} (DSR1)  and by
Magueijo and Smolin (DSR2)~\cite{ms}.  DSR has an explicit realization
in momentum space, in  the sense that the transformation laws for
the energy and momentum  of a particle in some reference frame are
known in terms of a generalized boost~\cite{ms,boos}.

The realization  of DSR in space-time is however a
subtle problem.  First, one notes that the  algebra that characterizes
DSR  is the  $\kappa$-Poincar\'e algebra  (KP) introduced  by Lukiersky,
Ruegg, Nowicki and Tolstoi in Ref.~\cite{luki}  and therefore, 
at the level of the algebras, 
different  DSR theories correspond to  different choices of  basis in
the  KP  context~\cite{kappadsr}.  The space-time  structure  can  be
introduced  in a  natural  way,  by defining  the  dual algebra  (Hopf
algebra)   of  KP.  The   so-called  $\kappa$-Minkowski (KM) 
space-time~\cite{kappamink} constructed in such a way has a 
non-trivial algebra.

But this construction of the space-time sector is not satisfactory at
all from  the DSR point of view. For example, if one  considers a
system of two particles, the four-momentum composition law inherited
from the co-product structure is not invariant under the
transformation laws of DSR~\cite{compo}.

On the  other  hand,  we  need  to  confront  such  proposals with
experiments  and  predictions.  DSR  phenomenology  has been  
studied~\cite{fenodsr} in  order to  test these new ideas from the kinematic
point  of view.  That  is, one starts  with  the deformed  dispersion
relation [for  DSR1, see Eq.~(\ref{casimir1})] plus   a  compatible
composition law, and studies the new features that DSR implies.

In this approach, processes involving particles and antiparticles are
very   interesting  because   the   assignment  of   the  energy   for
antiparticles is  not  a  trivial task.  In  the
standard case, the Casimir $E^2  -\boldsymbol{p}^2$ has two roots which are
symmetrical  under the  change $E\leftrightarrow  -E$ and, therefore,
there is no ambiguity in the interpretation of $E$ as the energy of
both the particle and the antiparticle. But in a DSR theory this 
interpretation is not so simple since in general there are 
two solutions for the energy, and they
turn out to be non-symmetric~\cite{fenodsr}.

In order to  solve this problem one should construct the quantum field
theory  compatible with DSR. However, a first indication  about the
particle-antiparticle content of the theory  can be traced back in the
Dirac equation (DE) compatible with DSR principles.

A proposal  for such operator  has been given in Ref.~\cite{DEamelino,alu} and
before it, in  the KP approach in Ref.~\cite{romp}. From both
definitions of the Dirac operator one can
extract information about the particle and antiparticle sectors and
also  contrast  them in  order to obtain  a  different perspective on
the relation between KM space-time and DSR theories.

In this paper we will explore the particle and antiparticle sectors
in DSR and  KM space-time.  In order to do that, we will first review
three different approaches in the LI case: a) by studying the solutions for the
energy in the dispersion relation; b) from the space-time solutions of
the  DE, and  c) from  the DE in momentum space. We will be able to
follow approaches a) and c) to analyze the antiparticle sector in DSR.
However, since we do not know how to write the DE compatible with DSR 
in space-time, we will study the KM deformed Dirac equation, which can 
also be written in momentum space. This will allow to compare the DSR
and KM results. 

The paper is organized as follows.  In Section~\ref{Sec:2} we will review
the LI case, along the lines described in the previous paragraph. 
Section~\ref{Sec:3} is devoted to the analysis of the DE in DSR1 and
KM space-time. Finally, in Section~\ref{Sec:4} the discussion and
conclusions are presented.

\section{Lorentz invariant particle-antiparticle sectors}
\label{Sec:2}

In  this section  we  will  review three  approaches  which permit  to
identify  the particle  and antiparticle  sectors in  the  standard LI
theory, namely:  the roots of  the energy in the  dispersion relation,
the DE in space-time, and the DE in momentum space.

These  approaches are very  well known  and, of  course, they  are not
enough to recognize and give a definite antiparticle interpretation,
which  becomes  clear  under  the  light of  a  quantum  field  theory
(QFT), but we will use them as a guide to identify antiparticles in DSR
theories since we do not have a proper QFT compatible with the DSR principle.

At the level of the Lorentz invariant theory, the DE in space-time
is enough to introduce the concept of antiparticle; however, we will
also consider the DE in momentum space in order to compare with DSR,
for which the connection between space-time and momentum formulations
is not known. 

\subsection{Dispersion Relation}
The first signal of the presence of antiparticles comes
from the dispersion relation itself
\begin{equation}
E^2 -\boldsymbol{p}^2=m^2,
\label{casimirli}
\end{equation}
which has two roots for the energy,
$E_\pm=\pm\sqrt{m^2+\boldsymbol{p}^2}$. The positive root $E_+$ 
corresponds to the particle energy, while $E_- $, being negative, appears to be
classically meaningless as an energy of a free particle.

In quantum field theory, however, the negative energy solutions
are connected with the existence of positive energy antiparticles,
with energy $|E_-|$, which therefore gives a physical meaning to the solution
coming from the negative sector. A useful picture to visualize the
concept of antiparticle is that of the ``Dirac sea''. Here the vacuum
consists of an infinitely deep sea of completely filled negative
energy levels. An antiparticle is associated to a ``hole'' in the sea,
that is, to a state in which all the negative energy levels but one
are occupied. The hole corresponds to the absence of a negative energy
$E_-$, and has therefore a larger energy than that of the vacuum,
exactly in the quantity $-E_-$. Since we assign to the vacuum zero energy and
momentum, the hole (antiparticle) associated to the absence of a
negative energy state of momentum $\boldsymbol{p}$ and energy $E_-$ satisfies
\begin{equation}
E_{antip}+E_-=0, \quad \boldsymbol{p}_{antip}+\boldsymbol{p}=0.
\label{sea}
\end{equation}
 
The point that we would like to emphasize is that the sector of negative
solutions  (their  modulus) in  Eq.~(\ref{casimirli})  has  a meaning  and
therefore, these solutions should not be discarded by hand. And the fact
that $E_+=|E_-|$ expresses a symmetry property, but it does not say that
 $E_+$ is  the energy of the antiparticle, it only says that both
energies coincide.

The role of both solutions becomes clearer from the Dirac equation,
which we will review briefly in the next subsection.

\subsection{Dirac equation in space-time} 
\label{Sec:DEst}

The Dirac equation in space-time is
\begin{equation}
\left(i\,
{\partial\hspace{-0.6em}\slash \hspace{0.15em}}-m\right)\psi(x)=0,
\label{deli}
\end{equation}
where${B \hspace{-0.6em}\slash
\hspace{0.15em}}=\gamma^\mu B_\mu $,  $\gamma^\mu$ are the
Dirac matrices and $m$ is the mass of the particle. 

The DE has two kind of linearly independent solutions
\begin{eqnarray}
\psi^{(+)}(x)&=&e^{- i p_\mu x^\mu}~u(p),\nonumber
\\
\psi^{(-)}(x)&=&e^{i p_\mu x^\mu}~v(p),
\label{desolli}
\end{eqnarray}
with\footnote{We are using the metric diag$(1,-1,-1,-1)$.} $p_0 >0$
and $u(p)$, $v(p)$ are bispinors which satisfy
\begin{eqnarray}
({p \hspace{-0.6em}\slash\hspace{0.15em}}-m)u(p)&=&0,\nonumber
\\
({p \hspace{-0.6em}\slash\hspace{0.15em}}+m)v(p)&=&0.
\label{deuvli}
\end{eqnarray}
In order to have non  trivial solutions for $u(p)$ and $v(p)$ the
condition~(\ref{casimirli}) must be satisfied. Note that the condition
$p_0>0$ implies that only positive roots must be considered.

Let us point out a fact that is trivial at this  level, but that will
be useful in the construction (and identification) of the antiparticle
sector in  DSR1. In Eq.~(\ref{deli}),
$\psi^{(+)}(x)$ is called the positive-frequency solution, and 
describes a particle of energy $p_0$ and momentum $\boldsymbol{p}$, while
$\psi^{(-)}(x)$ is named the negative-frequency solution, and corresponds to
a negative energy $-p_0$ and a momentum $-\boldsymbol{p}$. 
The interpretation of this
negative-frequency solution as an antiparticle follows the line of
arguments of the previous subsection. Defining the map
\begin{equation}
S_{\li}[(p^0,\boldsymbol{p})]= (-p^0,-\boldsymbol{p}),
\label{antipodeli}
\end{equation} 
we see that 
\begin{enumerate}
\item $S_{\li}$ maps the positive-frequency solution $\psi^{(+)}(x)$
into the negative-frequency solution $\psi^{(-)}(x)$, and
\item $S_{\li}$ maps the energy-momentum of the negative frequency
solution into the energy-momentum of the associated antiparticle state
[see Eq.~(\ref{sea})].
\end{enumerate}

This trivial map satisfies the following properties\footnote{We will
suppress the index $\mu$ in the  four vectors in order to simplify the
notation.}: a) $S_{\li}[S_{\li}[p]]=p$, \linebreak
b) $p + S_{\li}[p]=0$ and c) it leaves Eq.~(\ref{casimirli}) invariant. 

Note that properties a) and  b) lead to Eq.~(\ref{sea}) for the
energy-momentum of the antiparticle, 
and properties a) and c) are necessary
conditions for the two frequency solutions to be related by this
mapping.

In the  next subsection, we will  discuss another approach to the DE,
namely, the momentum  space formulation. We will of course arrive
to Eq.~(\ref{deuvli}), but without any explicit reference to space-time. 

\subsection{Dirac equation in the momentum space}
\label{Sec:DEmomentum}

We  are interested in  a DSR1
consistent approach;  therefore, it is necessary the discussion of the  
problem in momentum space, where DSR is well understood and properly defined.

In the LI case, the DE can be obtained by performing  a boost on, for
instance,  chiral  bi-spinors ($u_L(0),u_R(0)$)  defined  in the  rest
frame with the condition $u_L(0)= u_R(0)$~\cite{ryder}.

Under a boost with rapidity $\xi$, the spinors transform according to
\begin{eqnarray}
u_R(p)&=&[\cosh(\xi/2)+\boldsymbol{\sigma}\cdot\boldsymbol{n}\sinh(\xi/2)]
\,u_R(0),\nonumber \\ 
u_L(p)&=&[\cosh(\xi/2)-\boldsymbol{\sigma}\cdot\boldsymbol{n}\sinh(\xi/2)]
\,u_L(0),
\label{lidem}
\end{eqnarray}
where $\boldsymbol{n}$ is the direction of the momentum, $\sigma^i$
($i=1,2,3$) are the Pauli matrices, and $\xi$ the rapidity, defined by
$$
\cosh(\xi)=\frac{E}{m},\quad \quad\sinh(\xi)=\frac{p}{m}.
$$ 
The condition $u_L(0)= u_R(0)$ in Eq.~(\ref{lidem}) becomes
\begin{equation}
({p \hspace{-0.6em}\slash\hspace{0.15em}} -m)u(p)=0,
\label{deuli}
\end{equation}
that is, the DE for the particle sector, as discussed previously.

The  antiparticle  sector  can  be  obtained  from  here  by  applying
the  map $S_{\li}[p]$ discussed in the previous section, giving
\begin{equation}
({p \hspace{-0.6em}\slash\hspace{0.15em}}+m)v(p)=0,
\label{devli}
\end{equation}
where  we have  called  $v(p)=T  u(S_{\li}[p])$, and  $T$  is a  matrix
constructed from the $\gamma$ matrices.

\section{The particle-antiparticle problem in DSR and KM spacetime}
\label{Sec:3}

In the present section we apply the preceding arguments to the DSR
case. To do that,
we first discuss a general approach to DSR in momentum space, and 
then we obtain the map $S_{\dsr}$ analogous to the map $S_{\li}$ 
of the Lorentz invariant case, giving its explicit form for the DSR1 model. 

The map $S_{\dsr}$ will allow us to identify the antiparticle sector
just from the dispersion relation of a DSR theory. Taking the DSR1
model as a  specific example, we will then write the DSR1 DE in
momentum space, as it was obtained in Ref.~\cite{DEamelino}, and from
here we will be able to obtain the DE for the antiparticle.

The analysis of the DE in space-time compatible with the DSR
principles is not an easy task, as mentioned in  the
Introduction. When compared in momentum space, the DE in KM space-time
turns out to be the same as the DE in the DSR framework~\cite{DEamelino}.
However, the antiparticle  sector seems to be different in the two
approaches, as we will see below.

\subsection{DSR formulation}
\label{Sec:DSRfor}

It is convenient for our purposes to understand DSR  as a nonlinear
realization of the Lorentz group.  That is, we assume that there exists
a function $F:\Re^4\rightarrow\Re^4$ (with components $F^\mu$) which acts on momentum space
with  coordinates  $p^\mu=\{E,\boldsymbol{p}\}$, and maps it to the space
$\chi_{F}$   with  coordinates   $\pi^\mu=\{\epsilon,\boldsymbol{\pi}\}$ as
follows\footnote{The problem of identifying the original variables
  ($p^{\mu}$) as the physical ones in an operational (measurable) way is
  still an open problem.}: 
$$ \pi^\mu =\left(F^{-1}[p^\alpha]\right)^\mu, $$  
in  a  way  that  the  Lorentz group  acts  linearly  on  $\chi_F$
\cite{ms,chep,visser,unruh}.

This function allows, for instance, to construct the finite DSR boosts
($\mathcal{B}$) from the  standard Lorentz  boosts $\Lambda$ by
means of the composition of functions\footnote{We will suppress the index
on $F$ as we have done with $p$.}, 
$$ {\mathcal{B}}=F\circ\Lambda\circ F^{-1}. $$ 
The four-momentum composition law ($\hat{+}$) for two particles has
to satisfy two physical conditions:  $i$) it has to act linearly under
the  non-linear boosts  and $ii$)  it has  to be  invariant  under the
interchange of particle labels:
\begin{eqnarray}
\label{boostcompo}
\mathcal{B}[p_a~ \hat{+}~p_b]  &=&  \mathcal{B}[p_a]   ~\hat{+}  
~\mathcal{B}[p_b],\\ 
\mathcal{B}[p_a~ \hat{+}~p_b] &=&\mathcal{B}[p_b~ \hat{+}~p_a].
\end{eqnarray}
A natural definition  that satisfies both conditions is  given in terms
of the $F$ function:
\begin{equation}
p_a~\hat{+}~p_b=F\left[F^{-1}[p_a]+F^{-1}[p_b]\right],
\label{compodsr}
\end{equation}
which gives a DSR consistent law \cite{visser}.

In order to identify the antiparticle sector from the deformed DE, one
needs to construct a  map $S_{\dsr}[p]$, demanding the same properties
as in  the LI  case. However, one of these  properties is rather 
non-trivial in this case because it involves the composition law,  
namely $p + S_{li}[p]=0$, which  has  sense  only  if  it  is  valid  in  any  
reference  frame. The DSR compatible requirement will therefore be
$p\hat{+}S_{\dsr}[p]=0$.

It is direct to prove that the map
\begin{equation}
S_{\dsr}[p]=F\left[-F^{-1}[p]\right],
\label{anipoddsr} 
\end{equation}
satisfies all the  requirements, that is: a) $S_{\dsr}[S_{\dsr}[p]]=p$,
b)  $p\hat{+}S_{\dsr}[p]=0$,   and  c)  the   Casimir  element  remains
invariant.

For the DSR1 model, the explicit form of $F$ and its inverse is
\begin{eqnarray}
F[x,\boldsymbol{y}]&=&\left[\begin{array}{c}
\frac{1}{\lambda}\ln\left(\lambda
x+\sqrt{1+\lambda^2(x^2-\boldsymbol{y}^2)}\right)   \\   
\boldsymbol{y}\left[\lambda
x+\sqrt{1+\lambda^2(x^2-\boldsymbol{y}^2)}\right]
\end{array}\right],
\\                           
F^{-1}[x,\boldsymbol{y}]&=&\left[\begin{array}{c}
\frac{1}{\lambda}\left(\sinh(\lambda    x)+\frac{\lambda^2}{2}\boldsymbol{y}^2
e^{\lambda x}\right) \\ \boldsymbol{y}e^{\lambda x}
\end{array}\right].
\end{eqnarray}

And the map $S_{\dsr}[p]$ (by components) becomes
\begin{eqnarray}
S_{\dsr}^0[p]&=&-E+
\frac{1}{\lambda}\ln(1-\lambda^2\boldsymbol{p}^2e^{2\lambda E}), \nonumber  \\
\boldsymbol{S}_{\dsr}[p]&=&\frac{-\boldsymbol{p}~e^{2\lambda
E}}{1-\lambda^2\boldsymbol{p}^2e^{2\lambda E}}\,.
\label{maps}
\end{eqnarray}

It is interesting to note that this function $S_{\dsr}$ is the
image of $S_{\li}$,  defined in the space $\chi_F$, under the action
of   $F$.    That   is,   in   $\chi_F$,  the   map   connecting   the
particle-antiparticle   sectors  is   the   Lorentz  invariant   one:
$S_{\li}[\epsilon,\boldsymbol{\pi}]=(-\epsilon,-\boldsymbol{\pi})$.   
More precisely,
$$ S_{\dsr}= F\circ S_{\li}\circ F^{-1}.
$$

Now we are able to define the antiparticle sector in the DSR framework
and contrast it with the results of the KM approach. The next sections are
devoted to this discussion.

\subsection{Dispersion relation approach}

On a DSR1-type theory, the Casimir element can be written as
\begin{equation}
\frac{2}{\lambda^2}\cosh(\lambda E)-\boldsymbol{p}^2~e^{\lambda E}=
\frac{2}{\lambda^2}\cosh(\lambda m), 
\label{casimir1}
\end{equation}
where $\lambda$ is the invariant scale. $E$ and $\boldsymbol{p}$ are the
energy and momentum of the particle. The parameter $m$ is the rest
mass of the particle. 

The solutions of Eq.~(\ref{casimir1}) for the energy as a function of
momentum are 
\begin{equation}
E_{\pm} (\boldsymbol{p})=\frac{1}{\lambda}\ln\left(\frac{\cosh(\lambda
m)\pm\sqrt{\cosh^2(\lambda m)-(1-\lambda^2 \boldsymbol{p}^2)}}{1-\lambda^2
  \boldsymbol{p}^2}\right),
\label{edsr1}
\end{equation}
and the LI case is recovered in the limit $\lambda \rightarrow 0$,
where Eq.~(\ref{edsr1}) becomes $\pm \sqrt{\boldsymbol{p}^2 +m^2}$, 
that is, both sectors of energies. 

Note that in the rest frame ($|\boldsymbol{p}|=0$), Eq.~(\ref{edsr1}) becomes
\begin{equation}
E_{\pm} (0)=\pm m,      
\end{equation}
without any corrections due to DSR. The result does not depend on
$\lambda$ and therefore, the interpretation of $E_+$ as the energy of the
particle and $|E_-|$ as the energy of the antiparticle should
remain unchanged in this frame.

After a DSR1 boost, the energies of the particle and its
antiparticle can be read from Eq.~(\ref{edsr1}). The solution $E_-$ remains 
negative in any reference frame and this could suggest to associate
this solution with an antiparticle of energy $-|E_-|$. However, this
guess assumes implicitly that the energy and momentum of the
antiparticle are obtained from the negative-frequency solution
$(E_-(\boldsymbol{p}),-\boldsymbol{p})$ 
by applying the map $S_{\li}$. But this approach is
too naive: we should be consistent and use $S_{\dsr}$ instead. Then
the energy of the antiparticle becomes
\begin{eqnarray}
S_{\dsr}^0[E_-]&=&-E_-+\frac{1}{\lambda}\ln\left[e^{-\lambda E_-}\cosh(\lambda m)-1\right]\nonumber \\ &\sim&
    -E_--\lambda(E_-^2-m^2)-\mathcal{O}(\lambda^2).
\end{eqnarray}
\noindent From the previous equation we see that $S_{\dsr}^0[E_-]$ is positive
and, since the mapping $S_{\dsr}$ leaves the Casimir invariant, it
results  that 
$S_{\dsr}^0[E_-(\boldsymbol{p})]=E_+(\boldsymbol{S}_{\dsr}[\boldsymbol{p}])$.
On the other hand, since $E_-(\boldsymbol{p})$ does not depend on the
sign of $\boldsymbol{p}$, $S_{\dsr}^0[E_-(\boldsymbol{p})]=
S_{\dsr}^0[E_-(-\boldsymbol{p})]=E_+(\boldsymbol{S}_{\dsr}[-\boldsymbol{p}]).$

Therefore, the particle state $(E_+,\boldsymbol{p})$ is associated with the
antiparticle state
$(W,\boldsymbol{q})=(S_{\dsr}^0[E_-(\boldsymbol{p})],
\boldsymbol{S}_{\dsr}[-\boldsymbol{p}])$ where
$W=E_+(\boldsymbol{q})>0$. Both the particle and the antiparticle have 
the same relation between energy and momentum, i.e.~in both cases the 
dispersion relation ($C(E_+,\boldsymbol{p})=0$ and
$C(W,\boldsymbol{q})=0$) is defined by the same Casimir invariant
$C(p)$ and the energy dependence on the spatial
momentum is given by the positive solution of Eq.~(\ref{edsr1}). 
This is not a trivial result and it was crucial to use the properly 
defined $S_{\dsr}$ instead of the usual $S_{\li}$.

\subsection{DSR1 Dirac equation}

The DSR1  version of the DE can be  obtained in the same  way described in
Section~\ref{Sec:DEmomentum}, but now 
considering the DSR1 boost~\cite{DEamelino}
\begin{equation}
\cosh(\xi)=\frac{e^{\lambda E}-\cosh(\lambda m)}{\sinh(\lambda m)}.
\end{equation}
By using also the Casimir~(\ref{casimir1}), the DSR-DE becomes
\begin{equation}
\left({D     \hspace{-0.6em}\slash\hspace{0.15em}}     -~\sinh(\lambda
m)\right) u(p)=0,
\label{diracdsr1} 
\end{equation}
where
\begin{eqnarray}
D^\lambda_0&=&e^{\lambda         E}-\cosh(\lambda        m),        \\
D^\lambda_j&=&\lambda~e^{\lambda E}~p_j.
\label{diracdsr1oper}
\end{eqnarray}
Note that from the point of view of our approach to DSR 
(Section~\ref{Sec:DSRfor})
it is  possible to obtain the same equation~(\ref{diracdsr1}) just by
mapping  the DE from the  space  $\chi_F$  to  the  space  where  the
Lorentz  group  acts nonlinearly\footnote{One should take into
account that the parameter $m$ in the DSR1 theory is such that
$\frac{\sinh(\lambda m)}{\lambda}$ is the generalization of the
Lorentz invariant mass.}. In fact Eq.~(\ref{diracdsr1}) can be
rewritten as 
\begin{equation}
\left[\gamma^{\mu} F^{-1}_{\mu}[p] - \frac{\sinh(\lambda
    m)}{\lambda}\right] u(p) = 0.
\end{equation}

The antiparticle sector can be obtained as follows. From a
pure  DSR1 point of  view we  must consider the  map $S_{\dsr}[p]$
defined in Eq.~(\ref{maps}), under which the DE becomes
\begin{equation}
\left[\gamma^{\mu} F^{-1}_{\mu}\left[S_{\dsr}[p]\right] - \frac{\sinh(\lambda
    m)}{\lambda}\right] v(p) = 0.
\end{equation}
But taking into account that $S_{\dsr}= F\circ S_{\li}\circ F^{-1}$,
one has 
$$F^{-1}\left[S_{\dsr}[p]\right] = S_{li}\circ F^{-1}[p] = - F^{-1}[p]$$ 
and then the antiparticle spinor $v(p)$ satisfies the equation 
\begin{equation}
\left(\gamma^0\left[e^{\lambda        E}-\cosh(\lambda       m)\right]
+\gamma^ip_i~e^{\lambda E} +\sinh(\lambda m)\right)v(p)=0.
\label{dedsr1ap}
\end{equation}

Again, as  in the case of  particles, the DE for  antiparticles is the
image of Eq.~(\ref{deuvli}) in  the $\chi_F$ space under the  action of
$F$. 

In the next section we will  analyze the problem in KM space-time and
we will  see that  a  different map  is  needed in  order  to render  the
antiparticle sector.

\subsection{The   $\kappa$-Minkowski  space-time  approach}

In KM space-time,  things turn out to be  more complicated because we have
 non-commutativity in  the  sense\footnote{Note that  the  sign  of
$\lambda$ is  reversed if  we compare with  the original work  of Wess
\cite{wess}.  The reason  for doing that is two  fold: first, the sign
of $\lambda$  is not fixed  in the $\kappa$-Minkowski  deformation and
second,  in order to  match with  the results  given, for  instance in
Ref.~\cite{kow}.}
$$ [x_i,x_0]=i\lambda~x_i,
$$ with $\{i=1,2,3\}$, and $x_0$ the time coordinate.

In  order  to  find solutions  of  the  deformed  DE, we  will  follow
Ref.~\cite{wess}, using the $\star$-product formulation.  The  algebra  of
functions is given by
\begin{equation}
\left(f\star g\right)(x)=f(x)g(x)-\frac{i\lambda}{2}(x^\ell\partial_\ell
f(x)\partial_0 g(x)-x^\ell\partial_\ell g(x)\partial_0 f(x))\dots,
\label{star}
\end{equation} 
with $\lambda$ the DSR invariant scale.
	
The derivatives  in the non-commutative  space-time $\hat\partial_\mu$
are mapped  to standard derivatives  in the commutative  space-time as
follows:
\begin{eqnarray}
\hat{\partial}_0&\longrightarrow      \partial^*_0&=\partial_0,     \\
\hat{\partial}_j&\longrightarrow
\partial^*_j&=\frac{e^{-i\lambda\partial_0}-1
}{-i\lambda\partial_0}\partial_j.
\label{deriva}
\end{eqnarray}

The  Dirac   operator  (DO)  is  constructed   in  the  non-commutative
space-time by demanding  that it  transforms  as a  vector, and  the
requirement that it has the standard DO limit when $\lambda\rightarrow
0$. The image of the DO in the commutative space turns out to be
\begin{eqnarray}
D^*_0&=&\frac{1}{\lambda}\sin(\lambda\partial_0)-\frac{i\nabla^2}{\lambda
\partial_0^2}\left(\cos(\lambda\partial_0)-1\right),                 \\
D^*_j&=&\frac{e^{i\lambda\partial_0}-1}{i\lambda\partial_0}\partial_j,
\label{camposdsr}
\end{eqnarray}
where $\nabla^2=\partial_j~\partial_j$.

In  order   to  find   solutions  of  the   DE,  let  us   define  the
four-momentum-like operator $\Pi_\mu$, on the commuting space-time
\begin{eqnarray}
\Pi_j&=&\frac{1-e^{-i\lambda\partial_0}}{-\lambda\partial_0}~\partial_j,
\nonumber \\ \Pi_0&=&i\partial_0.
\label{momentum}
\end{eqnarray}
One can verify that the class of functions
\begin{equation}
\phi^{(+)}(x)=e^{-i\left[E\,t-i\left(\frac{\lambda E}{1-e^{-\lambda
E}}\right)\boldsymbol{p}\cdot\boldsymbol{x}\right]},
\label{onda}
\end{equation}
satisfies
\begin{eqnarray}
\Pi_\mu~\phi^{(+)}(x)&=&p_\mu~\phi^{(+)}(x),
\label{eigen}
\\                           \phi^{(+)}(x)&\stackrel{\lambda\rightarrow
0}{\longrightarrow}&e^{-i~p_\mu x^\mu},
\label{limli+}
\end{eqnarray}
with $p^\mu=(E,\boldsymbol{p})$. The last equation suggests to interpret this
function as the one corresponding to the particle sector solutions.

It is also straightforward to prove that this class of solutions also
satisfies
\begin{eqnarray}
D_0^*~\phi^{(+)}(x)&=&-\frac{i}{\lambda}\left[e^{\lambda
E}-\cosh(\lambda m) \right]~\phi^{(+)}(x),         \\
D_i^*~\phi^{(+)}(x)&=&-i~e^{\lambda E}p_i~\phi^{(+)}(x),
\end{eqnarray}
where we have used the $\kappa$-Poincar\'e dispersion relation which is
the same as Eq.~(\ref{casimir1})~\cite{wess}.

Therefore, the function
\begin{equation}
\psi(x)=\phi^{(+)}(x)~u(p),
\label{solutionkapau}
\end{equation}
with $\phi^{(+)}(x)$ given in Eq.~(\ref{onda}), is a solution of the DE
\begin{equation}
\left(i\,{D^* \hspace{-1em}\slash\hspace{0.15em}}
-\lambda^{-1}\sinh(\lambda m)\right)\psi(x)=0,
\end{equation}
provided that $u(p)$ is a solution of
\begin{equation}
\left\{\left[e^{\lambda     E}-\cosh(\lambda     m)    \right]\gamma^0
+\lambda~e^{\lambda E}~p_i\gamma^i-\sinh(\lambda m)\right\}u(p)=0.
\label{dekmu}
\end{equation}

It is clear  that the condition to have a  non-trivial solution for $u$
is  the dispersion  relation given  by Eq.~(\ref{casimir1}).   Finally, we
conclude that  the energy  of the particle  sector corresponds  to the
positive root of the deformed dispersion relation, as we expected. 

As was noted in Ref.~\cite{DEamelino},  we also see that the momentum space
equation (\ref{dekmu}),  derived from KM space-time,  coincides with
the DSR1 Dirac equation~(\ref{diracdsr1}).

In order to find the other  sector, the antiparticle one, we must look
for   the   solutions   with   the  limit   $e^{ip_\mu   x^\mu}$   for
$\lambda\rightarrow 0$. In doing this we will proceed as in the previous
cases using the $S$ mapping as a natural tool. 

In defining $\skp$ for KP we first need the composition
law  for  four-momentum.  This  can  be  derived  from  the  coproduct
structure of the KP model and has the peculiarity to be non-symmetric:
\begin{equation}
\label{sumaKP}
E_a  \dot{+}  E_b  =  E_a  +  E_b \quad \quad \boldsymbol{p}_a
\dot{+}\boldsymbol{p}_b=\boldsymbol{p}_a +e^{\lambda E_a}\boldsymbol{p}_b.
\end{equation}
In  principle  this  gives  the  possibility  to  have  two  different
definitions of  $\skp $, i.e.
$$ p_a\dot{+}\skp[p_a]=0 \qquad \qquad \skp[p_a]\dot{+}p_a=0.
$$ 
It is remarkable that both definitions give the same result
\begin{eqnarray}
S_{\mbox{\small{{\it kp}}}}^0~[p]&=&-E,\\
\boldsymbol{S}_{\mbox{\small{{\it
	kp}}}}~[p]&=&-\boldsymbol{p}~e^{\lambda E}. 
\end{eqnarray}
This  map  is well  known  in  the context  of  Hopf  algebras and  it
corresponds to the antipodal mapping. 

As it was shown  in the LI case and  used  in the DSR1 model, in
this case we will demand again that the antiparticle four-momentum 
$(W,\boldsymbol{q})$ be related to the four-momentum of the negative frequency
solution $p\equiv (E_-,\boldsymbol{p})$ by
$W=S_{\mbox{\small{{\it kp}}}}^0~[p] $ and 
$\boldsymbol{q}=\boldsymbol{S}_{\mbox{\small{{\it kp}}}}~[p]$. 

As a consequence of the general properties defining the $S$ mapping
one has that, although the relation between the energy-momentum
of the negative frequency solution and the energy-momentum of the
antiparticle differs from the DSR1 case, once more the energy of the
antiparticle depends on the momentum in the same way as the energy of
the particle.

The  property $S[S[p]]=p$  allows to  write $p$  in terms  of  $q$ and
obtain the solution of the DE for the antiparticle as follows. The function
\begin{equation}
\phi^{(-)}(x)=e^{i\left[W\,t-\left(\frac{\lambda  W}{1-e^{\lambda
W}}\right)\boldsymbol{q}\cdot\boldsymbol{x}\right]},
\label{onda-}
\end{equation}
satisfies
\begin{eqnarray}
\Pi_\mu~\phi^{(-)}(x)&=&-q_\mu~\phi^{(-)}(x),
\label{eigen-}
\\ 
\phi^{(-)}(x)&\stackrel{\lambda\rightarrow
0}{\longrightarrow}&e^{i~q_\mu x^\mu}.
\label{limli-}
\end{eqnarray}
It is direct to prove that the function
\begin{equation}
\psi(x)=\phi^{(-)}(x)~v(p),
\label{solutionkapav}
\end{equation}
with $\phi^{(-)}(x)$ given by Eq.~(\ref{onda-}), is a solution of the DE
\begin{equation}
\left(i\,{D^*  \hspace{-1em}\slash\hspace{0.15em}}
-\lambda^{-1}\sinh(\lambda m)\right)\psi(x)=0,
\end{equation}
provided that $v(p)$ is a solution of
\begin{equation}
\left\{\left[e^{-\lambda     w}-\cosh(\lambda    m)    \right]\gamma^0
-\lambda~p_i\gamma^i-\sinh(\lambda m)\right\}v(q)=0.
\label{dekmv}
\end{equation}
The  DE for the antiparticle spinor  is different from
the one  obtained in  the DSR1  model. This is  due to  the different
composition laws, which induce different $S$ mappings. 

\section{Conclusions and Discussion}
\label{Sec:4}

In this  paper we have studied the  antiparticle sector in  the context of
the  DSR1  proposal and  the  $\kappa$-Minkowski space-time,  following
three different approaches inspired  by the Lorentz invariant case: a)
the dispersion relation,  b) the Dirac equation in space-time and
c) the Dirac equation in momentum space.

For the DSR proposal, whose consistent realization in space-time is
not known, just a) and c)  are well defined.  On the other hand, there
is  evidence  that, at  least  in the  one  particle  sector, the  KM
approach could  be compatible with the DSR principles and  therefore, the
test b) could give another insight on such relation.

In Section~\ref{Sec:2}, in a  LI analysis, it is possible to realize that
the  sector of  particles and  antiparticles  are connected  by a  map
($S_{\li}$),  which is  trivial in  a LI  theory, with  the properties
pointed out in  Section~\ref{Sec:DEst}. The interesting point  is that one of
such  properties requires  the  composition law:  a  rule for  summing
momentum that must be covariant. When  this law is the standard sum of
four vectors, the covariance is guaranteed by the linear nature of 
Lorentz transformations.

>From a pure DSR point of view, we showed that it is possible to define
a map $S_{\dsr}$  which gives the antiparticle sector.  Such map turns
out to be  the image of $S_{\li}$ by the function  $F$ in the approach
of \cite{ms,visser,unruh} to DSR.

Then, following  a), we  showed  that for  DSR1 one  should take  the
positive  root  of  the  dispersion  relation as  the  energy  of  the
antiparticle.  The  negative solution  has not  a  physical meaning;
however, when  it is  mapped by $S_{\dsr}$  it renders  the antiparticle
energy.  But because  the spatial  momentum is  also mapped  in  a 
non-trivial way, the energy of the antiparticle depends on its spatial
momentum in the same way as  the energy of the particle depends on its
momentum, and both are positive\footnote{In fact in the case of a
  spinless system, the identification of the $S$ map at the level of the
  dispersion relation is all one needs to establish the relation
  between the particle and antiparticle sectors.}. This result is
consistent with some phenomenological approaches~\cite{fenodsr,fenonl}.

In the approach c), from the  DE in momentum space (as was obtained
in Ref.~\cite{DEamelino},  for   instance),  the  antiparticle  sector  is
straightforwardly obtained through the  map $S_{\dsr}$.  We found that
the DE for the antiparticle is the image under $F$ of the standard DE
(for antiparticles) in the space $\chi_F$.

>From  the   space-time  point  of   view  [approach  b)], it is
possible to find a solution of the deformed DE in  
KM  space-time which, in momentum space,  is the  same as  the DE
compatible with DSR1.

However,  here  we  can not  use  the  map  $S_{\dsr}$ to  obtain  the
antiparticle sector. In fact, if we  look for a function $S$ with the
properties discussed  in Section~\ref{Sec:DEst}, there is a  
crucial difference,
that is, the rule for summing  momenta is not $\hat{+}$, the DSR one;
here  instead, such  rule is  given by  the coproduct  structure. This
fixes the  map $S$, (which turns out  to be the antipodal  map, in the
Hopf algebra language) and because of that the antiparticle sector in
KP is different from that of DSR1. 

This difference is expressed in the Dirac equation. The deformed DE
satisfied by  a DSR1  antiparticle is different  from the  deformed DE
satisfied by a KM particle.  The dispersion relations are the same, but
clearly their physical content is rather different. This  physical
content,  at the  end,  is  codified in  the composition
law which is crucial for any physical process.

Let  us comment  that  our result  can  be  viewed as  another
argument  showing  the  independence  between  DSR  and  KM theories (and
$\kappa$-Poincar\'e). However,  it clearly  shows that the  ideas coming
from the last one might give some light in the construction of a DSR in
space-time.

As a final remark, the identification of the $S$ map between
particle and antiparticle sectors suggests a generalization, 
$$\Psi(x) = \int
\frac{d^3\boldsymbol{F}[p]}{(2\pi)^3}\frac{1}{2F^{-1}_0[p]} 
\left[a_p u(p) e^{-ip.x} + b^{\dagger}_p v(p)
  e^{-iS[p].x}\right],$$ 
 of the relativistic Dirac quantum field of a Lorentz invariant theory
 as an starting point for a quantum field formulation of a
 DSR theory.  

\acknowledgments

This work has been partially supported by an INFN-CICyT collaboration
and MCYT (Spain), grant FPA2003-02948. 
We would like to thanks  G. Amelino-Camelia, J. Gamboa, J. Lukierski and A. Nowicki
for discussions. This work was partially developed during the $40^{th}$ 
Winter School on Theoretical  Physics, {\em Quantum gravity phenomenology}, 
 4-14 February (2004), Ladek  Zdroj, Poland. F. M. would like also to
 thanks the INFN for a  postdoc fellowship.

\end{document}